\documentclass[useAMS,usenatbib]{mn2e}
\usepackage{multirow}
\usepackage{epsfig}

\makeatletter
\let\jnl@style=\rm
\def\ref@jnl#1{{\jnl@style#1}}

\def\aj{\ref@jnl{AJ}}                   
\def\araa{\ref@jnl{ARA\&A}}             
\def\apj{\ref@jnl{ApJ}}                 
\def\apjl{\ref@jnl{ApJ}}                
\def\apjs{\ref@jnl{ApJS}}               
\def\ao{\ref@jnl{Appl.~Opt.}}           
\def\apss{\ref@jnl{Ap\&SS}}             
\def\aap{\ref@jnl{A\&A}}                
\def\aapr{\ref@jnl{A\&A~Rev.}}          
\def\aaps{\ref@jnl{A\&AS}}              
\def\azh{\ref@jnl{AZh}}                 
\def\baas{\ref@jnl{BAAS}}               
\def\jrasc{\ref@jnl{JRASC}}             
\def\memras{\ref@jnl{MmRAS}}            
\def\mnras{\ref@jnl{MNRAS}}             
\def\pra{\ref@jnl{Phys.~Rev.~A}}        
\def\prb{\ref@jnl{Phys.~Rev.~B}}        
\def\prc{\ref@jnl{Phys.~Rev.~C}}        
\def\prd{\ref@jnl{Phys.~Rev.~D}}        
\def\pre{\ref@jnl{Phys.~Rev.~E}}        
\def\prl{\ref@jnl{Phys.~Rev.~Lett.}}    
\def\pasp{\ref@jnl{PASP}}               
\def\pasj{\ref@jnl{PASJ}}               
\def\qjras{\ref@jnl{QJRAS}}             
\def\skytel{\ref@jnl{S\&T}}             
\def\solphys{\ref@jnl{Sol.~Phys.}}      
\def\sovast{\ref@jnl{Soviet~Ast.}}      
\def\ssr{\ref@jnl{Space~Sci.~Rev.}}     
\def\zap{\ref@jnl{ZAp}}                 
\def\nat{\ref@jnl{Nature}}              
\def\iaucirc{\ref@jnl{IAU~Circ.}}       
\def\aplett{\ref@jnl{Astrophys.~Lett.}} 
\def\apspr{\ref@jnl{Astrophys.~Space~Phys.~Res.}}
\def\bain{\ref@jnl{Bull.~Astron.~Inst.~Netherlands}}
\def\fcp{\ref@jnl{Fund.~Cosmic~Phys.}}  
\def\gca{\ref@jnl{Geochim.~Cosmochim.~Acta}}   
\def\grl{\ref@jnl{Geophys.~Res.~Lett.}} 
\def\jcp{\ref@jnl{J.~Chem.~Phys.}}      
\def\jgr{\ref@jnl{J.~Geophys.~Res.}}    
\def\jqsrt{\ref@jnl{J.~Quant.~Spec.~Radiat.~Transf.}}
\def\memsai{\ref@jnl{Mem.~Soc.~Astron.~Italiana}}
\def\nphysa{\ref@jnl{Nucl.~Phys.~A}}   
\def\physrep{\ref@jnl{Phys.~Rep.}}   
\def\physscr{\ref@jnl{Phys.~Scr}}   
\def\planss{\ref@jnl{Planet.~Space~Sci.}}   
\def\procspie{\ref@jnl{Proc.~SPIE}}   

\makeatother

\title[{Fe\,XXV} and {Fe\,XXVI} lines from low velocity, photoionised gas in the X-ray spectra of AGN]{{Fe\,XXV} and {Fe\,XXVI} lines from low velocity, photoionised gas in the X-ray spectra of AGN}
\author[Stefano Bianchi, Giorgio Matt, Fabrizio Nicastro, Delphine Porquet and Jacques Dubau]{Stefano Bianchi,$^{1}$\thanks{E-mail:
bianchi@fis.uniroma3.it (SB)} Giorgio Matt,$^{1}$ Fabrizio Nicastro,$^{2}$ Delphine Porquet$^{3}$ 
\newauthor
and Jacques Dubau$^4$\\
$^{1}$Dipartimento di Fisica, Universit\`a degli Studi Roma Tre, Via della Vasca Navale 84, I-00146, Roma, Italy\\
$^{2}$Harvard-Smithsonian Center for Astrophysics, Cambridge, MA, USA\\
$^{3}$Max-Planck-Institut f\"{u}r extraterrestrische Physik, Postfach 1312, 85741 Garching, Germany\\
$^{4}$LIXAM, Universit\'e Paris-Sud, 91405 Orsay cedex, France}

\begin{document}

\pagerange{\pageref{firstpage}--\pageref{lastpage}} \pubyear{2004}

\maketitle

\label{firstpage}

\begin{abstract}
  
  We have calculated the equivalent widths of the absorption lines
  produced by {Fe\,\textsc{xxv}} and {Fe\,\textsc{xxvi}} in a
  Compton-thin, low-velocity photoionised material illuminated by the
  nuclear continuum in AGN. The results, plotted against the
  ionisation parameter and the column density of the gas, are a
  complement to those presented by \citet{bm02} for the emission lines
  from the same ionic species. As an extension to the work by
  \citet{bm02}, we also present a qualitative discussion on the
  different contributions to the He-like iron emission line complex in
  the regimes where recombination or resonant scattering dominates,
  providing a useful diagnostic tool to measure the column density of
  the gas. Future high resolution missions (e.g., \textit{Astro-E2})
  will allow us to fully take advantage of these plasma diagnostics.
  In the meantime, we compare our results with an up-to-date list of
  Compton-thick and unobscured (at least at the iron line energy)
  Seyfert galaxies with emission and/or absorption lines from H- and
  He-like iron observed with \textit{Chandra} and XMM-\textit{Newton}.

\end{abstract}


\section{Introduction}

The present generation of X-ray missions, XMM-\textit{Newton} and \textit{Chandra}, are revealing new features in the spectra of AGN. An interesting case is represented by the observation of emission lines from {Fe\,\textsc{xxv}} and {Fe\,\textsc{xxvi}} in unobscured Seyfert galaxies \citep[see e.g.][ for the incidence of these features in a sample of bright objects]{bianchi04}. \citet{bm02} have shown that these narrow lines can share a common origin with those more commonly found in obscured objects, likely produced in a Compton-thin, photoionised material illuminated by the nuclear continuum \citep[see e.g.][ for one of the best studied example, NGC~1068]{matt04}.

This interpretation is in agreement with orientation-based unification models of AGN \citep[like the archetypal one introduced by][]{antonucci93}. Indeed, the same photoionised material should be present both in obscured and in unobscured objects, the only difference being that in the latter its presence is more difficult to observe because the lines are diluted by the nuclear continuum.

Another piece of information was added to this scenario very recently, with the detection of ionised iron lines also in absorption. Though in some cases the line energies are so blueshifted to imply high velocity outflows \citep[e.g.][and references therein]{reev03,pounds03b}, there is now significant evidence of {Fe\,\textsc{xxv}} absorption lines occurring in material with low velocity with respect to the rest frame of the source \citep{kaspi02,matt04b,vf04,reev04}. In the latter situation, the photoionised matter is likely the same revealed in emission in other objects but observed along the line of sight.

The plots of \citet{bm02} were produced by summing up all the transitions of a given ion. Although this is a good approximation for such weak emission lines when observed with the CCD resolution of the instruments aboard XMM-\textit{Newton} or the limited effective area available at the iron energy with \textit{Chandra}-HETGS (but see below for some possible diagnostics even in these cases), it will be no longer sufficient with future, high energy resolution missions, such as \textit{Astro-E2}.
With an energy resolution of 6 eV (FWHM) and an effective area of 150 cm$^{-2}$ at 6 keV, the calorimeters aboard Astro-E2 will be capable of resolving both the \mbox{Fe\,\textsc{xxvi}} K$\alpha$ doublet and the \textit{w}, \textit{x}, \textit{y} and \textit{z} lines from \mbox{Fe\,\textsc{xxv}}.

In this paper, we investigate the properties of the photoionised material as observed through absorption and emission features from {Fe\,\textsc{xxv}} and {Fe\,\textsc{xxvi}}. Following the approach by \citet{bm02} for the emission lines, we calculate the equivalent widths (EWs) of the absorption lines as a function of the ionisation parameter and the column density of the intervening material, taking advantage of the code discussed by \citet{nfm99}. As for the emission lines, while we refer the reader to \citet{bm02} for detailed calculations and plots, we devote Sect. \ref{emi} to a qualitative discussion on the relative contributions of resonance, intercombination and forbidden transitions to the total EWs for {Fe\,\textsc{xxv}}. Subsequently, we compare the theoretical results to an up-to-date list of observational evidence in favour of the presence of these features, in order to derive some physical parameters on the photoionised material responsible for their production.

\section{Theoretical models}

Table \ref{parameters} shows all the lines considered in this paper. The \mbox{Fe\,\textsc{xxv}} K$\alpha$ complex includes four lines: three corresponding to the transition $1s^2 - 1s2p$, subdivided into the resonance line $w$ ($^1S-^1P$) and the two intercombination lines $x$ ($^1S-^3P_2$) and $y$ ($^1S-^3P_1$), and one to $1s^2 - 1s2s$, that is the forbidden line $z$ ($^1S-^3S$). On the other hand, the \mbox{Fe\,\textsc{xxvi}} K$\alpha$ line is composed of a resonance doublet ($1s-2p$), similarly to neutral iron: the K$\alpha_1$ ($^2S_{1/2}-^2P_{3/2}$) and the K$\alpha_2$ ($^2S_{1/2}-^2P_{1/2}$). For  \mbox{Fe\,\textsc{xxvi}}, the ratio 2:1 of the oscillator strengths for K$\alpha_1$ and K$\alpha_2$ is fixed because it depends only on the statistical weights of the two upper levels. On the other hand, the behaviour of the four He-like transitions is far more complex, depending on several physical parameters of the matter, and deserves a detailed, though qualitative, discussion. Satellite lines are \textit{not} included in our calculations. Their possible relevance is discussed in Sect. \ref{satellite}.

\subsection{\label{abs}Absorption lines}

We consider the absorption features due to resonance transitions from {Fe\,\textsc{xxv}} and {Fe\,\textsc{xxvi}}, including the intercombination \textit{y} line for the former (see Table \ref{parameters} for all the details). On the other hand, we do not consider in our calculations the He-like iron intercombination \textit{x} and forbidden \textit{z} lines, because their very low oscillator strengths make them negligible in absorption.

\begin{table}
\caption{\label{parameters}The adopted parameters for the theoretical models developed in this paper (see text for details). All line parameters are taken from the NIST Atomic Database (http://physics.nist.gov/cgi-bin/AtData/main\_asd and references therein), if not otherwise stated.}

\begin{tabular}{ccccccc}
\hline
\multicolumn{7}{c}{\textit{Continuum}}\\
\hline
\multicolumn{3}{c}{\textsc{Table power law $\Gamma$}} & \multicolumn{4}{c}{2.0}\\
\multicolumn{3}{c}{\textsc{$n_e$}} & \multicolumn{4}{c}{$10^6$ cm$^{-3}$}\\
\multicolumn{3}{c}{\textsc{T}} & \multicolumn{4}{c}{$10^6$ K}\\
\multicolumn{3}{c}{$A_{Fe}$} & \multicolumn{4}{c}{4.68$\times10^{-5}$ $^a$}\\
\hline
\textit{Ion} & $E_m$ $^b$ &\textit{Line Id.} & \textit{E} & $f_{lu}$ & $g_i$ & $g_k$\\
&\textit{(keV)}&&\textit{(keV)}&&&\\
\hline
\multirow{4}*{\mbox{Fe\,\textsc{xxv}}} & \multirow{4}*{6.697}&K$\alpha$ \textit{w} (r) & 6.700 & 0.704 & 1 & 3\\
& &K$\alpha$ \textit{x} (i) & 6.682 & $1.7\times10^{-5}$ & 1 & 5\\
& &K$\alpha$ \textit{y} (i) & 6.668 & 0.069 & 1 & 3\\
& &K$\alpha$ \textit{z} (f) & 6.637 & $3.3\times10^{-7}$ & 1 & 3\\
\multirow{2}*{\mbox{Fe\,\textsc{xxvi}}} & \multirow{2}*{6.966} &K$\alpha_1$ (r) & 6.973 & 0.277$^c$ & 2 & 4 \\
& &K$\alpha_2$ (r) & 6.952 & 0.139$^c$ & 2 & 2 \\
\hline\\
\end{tabular}

$^a$ With respect to H \citep{ag89}

$^b$ Mean energy for the absorption lines considered in this paper, weighted on the oscillator strengths of the transitions contributing to the blend

$^c$ From \citet{vvf96}.

\end{table}

The equivalent width of an absorption line can be expressed as:
\begin{equation} \label{ew}
\mbox{EW} = \int_0^{+\infty} \left[ 1 - e^{-\tau_{\nu}} \right] d\nu,
\end{equation}
where $\tau_{\nu}$ is the dimensionless frequency-specific optical depth of the considered transition:
\begin{equation} \label{taunu}
\tau_{\nu} = \int_0^L ds \alpha_{\nu} = n_l L {{\pi e^2} \over {m_e c}}
f_{lu} \Phi(\nu).
\end{equation}

In this equation, $n_l$ is the number density of the relevant ion in the lower level, while $\Phi(\nu)$ is the normalised Voigt profile, which basically depends on the natural width of the considered transition and the Doppler width of the line, which in turn is a function of the gas temperature and its turbulence, that is its dispersion velocity along the line of sight.

The relative abundances of the {Fe\,\textsc{xxv}} and {Fe\,\textsc{xxvi}} ions are calculated by means of the photoionisation code \textsc{Cloudy} \citep[version 94.00: ][]{ferl00}, as a function of the ionisation parameter and the column density. Table \ref{parameters} summarizes the input parameters chosen to model the continuum in our calculations, where not otherwise stated.

Following \citet{bm02}, we adopted a modified ionisation parameter \citep[similar to the one first proposed by][]{netzer96}, which is more sensitive to the number of X-ray photons, so that our results are less dependent on the assumed continuum shape. The relation between this X-ray ionization parameter and the one adopted in \textsc{Cloudy} is a function of the power law index $\Gamma$ \citep{bm02}:

\begin{equation}
U_{x}=\frac {\int _{2}^{10}\frac{L_{E}}{E} dE} {4\pi r^{2}cn_{e}} \: \Longrightarrow \: \frac {U_{x}} {U} (\Gamma) =\frac { 2^{1-\Gamma} - 10^{1-\Gamma}} {E_{R}^{1-\Gamma}}
\end{equation}
where $E_{R}$ is the energy equivalent in keV of 1 Rydberg. 

The integration of the Voigt profile in the different column density regimes was performed following the methods and the code developed by \citet{nfm99}, to which we refer the reader for details. All the calculations were performed separately for the four transitions considered in this paper, while only their final EWs were summed when stated. It should be noted that this paper explores also large column densities, where another useful diagnostic tool for this gas can be represented by the depth of the photoelectric edges, at $\simeq8.8$ (\mbox{Fe\,\textsc{xxv}}) and $\simeq9.3$ (\mbox{Fe\,\textsc{xxvi}}) keV. However, a full treatment of these edges is beyond the scopes of this work.

\begin{figure}

\epsfig{file=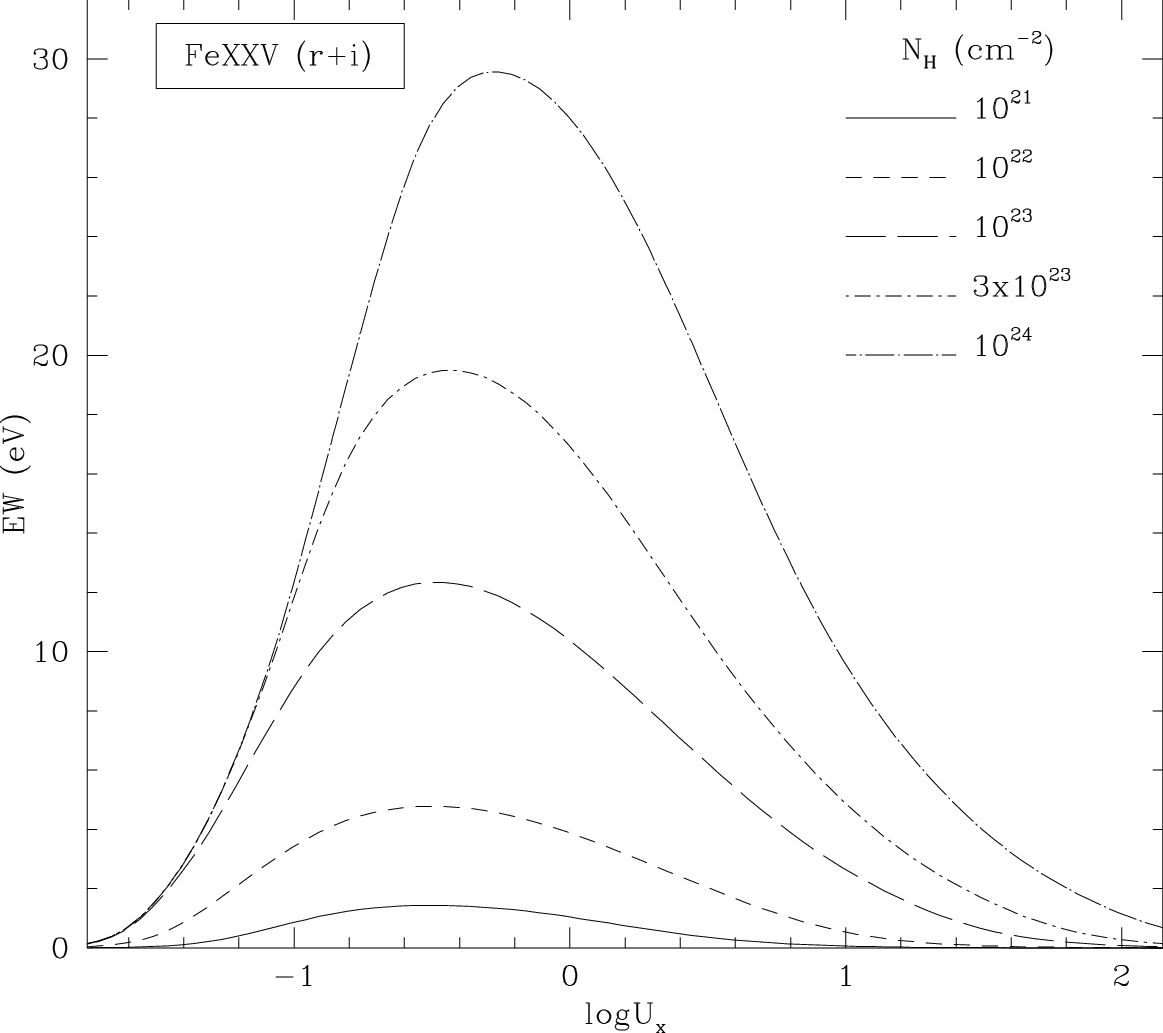, width=8cm}

\caption{\label{fe25abs}EW for the Fe\,\textsc{xxv} K$\alpha$ absorption complex (resonant + intercombination) as a function of $U_x$, for different values of column densities. See Table \ref{parameters} for the parameters adopted in the calculations. Turbulent velocity is set equal to zero.}

\end{figure}

Figures \ref{fe25abs} and \ref{fe26abs} show the expected EWs for \mbox{Fe\,\textsc{xxv}} (resonant plus intercombination) and \mbox{Fe\,\textsc{xxvi}} (K$\alpha_1$ plus K$\alpha_2$) as a function of $U_x$ and for different column densities. In these plots, the turbulence is assumed to be null. Figure \ref{ratio} shows the ratio between the EWs from the two ions, as a function of $U_x$. Indeed, the \mbox{Fe\,\textsc{xxv}}/ \mbox{Fe\,\textsc{xxvi}} ratio, being not sensitive to the column density in the most interesting range of ionisation (i.e. around $\log U_x=0$ where the EWs of \mbox{Fe\,\textsc{xxv}} and \mbox{Fe\,\textsc{xxvi}} are equal as shown in Figures \ref{fe25abs} and \ref{fe26abs}), can be used to measure the latter from the data.

\begin{figure}

\epsfig{file=ME1051rv_2.eps, width=8cm}

\caption{\label{fe26abs}EW for the Fe\,\textsc{xxvi} K$\alpha$ absorption complex (K$\alpha_1$ + K$\alpha_2$) as a function of $U_x$, for different values of column densities. See Table \ref{parameters} for the parameters adopted in the calculations. Turbulent velocity is set equal to zero.}

\end{figure}

\begin{figure}

\epsfig{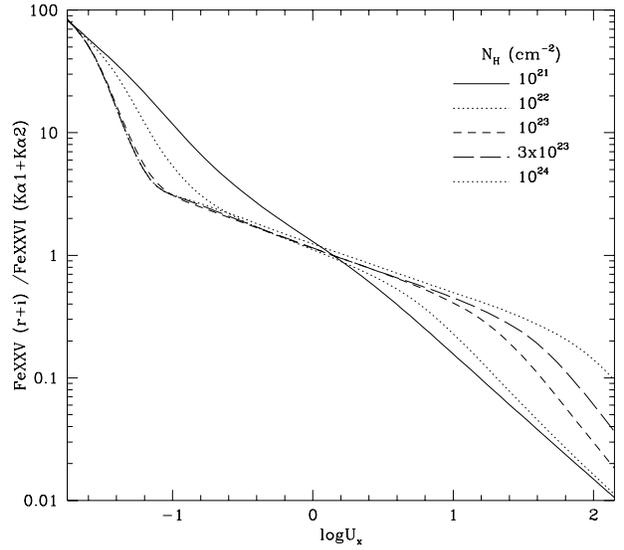}

\caption{\label{ratio}Ratio of the absorption lines EWs of the Fe\,\textsc{xxv} K$\alpha$ absorption complex (resonant + intercombination) over Fe\,\textsc{xxvi} (K$\alpha_1$ + K$\alpha_2$) as a function of $U_x$, for different values of column densities. See Table \ref{parameters} for the parameters adopted in the calculations. Turbulent velocity is set equal to zero.}

\end{figure}

\begin{figure}

\epsfig{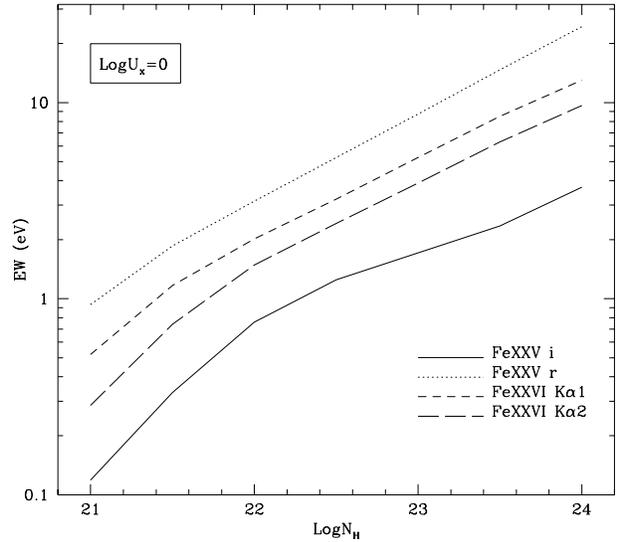}

\caption{\label{cog}Curve of growth for the four transitions considered in this paper, with $\log U_x=0$. See Table \ref{parameters} for the parameters adopted in the calculations. Turbulent velocity is set equal to zero.}

\end{figure}

The curve of growths for each transition separately are plotted in figure \ref{cog} \footnote{See Nicastro et al., in preparation, Errata, for an explanation of the difference between the curve of growth for Fe\,\textsc{xxvi} plotted here and that plotted in Fig. 1 of the original work by \citet{nfm99}}. These curves, calculated at $\log U_x=0$, give a feeling of the different contributions of the transitions to the total EW. Note that, for low column densities, their ratios are simply dictated by the ratios of their oscillator strengths, while some deviations, though quite marginal, can be observed at larger column densities.

We now consider the effects of the turbulence. At a temperature of T=$10^6$ K, the mean thermal velocity of iron ions is generally expressed by the Doppler parameter:

\begin{equation}
b\equiv \sqrt{\frac{2\,kT}{M}}\simeq 17 \, \mathrm{km\, s^{-1}}
\end{equation}

which means that, for any velocity dispersion larger than $\sigma_v\simeq100$ km s$^{-1}$, the thermal motion will contribute less than 15\% to the total Doppler broadening. Note that turbulent velocities of that order are easily reached in a gas in Keplerian motion at a parsec scale around a SuperMassive BH, so that they should be included in a realistic model.

\begin{figure}

\epsfig{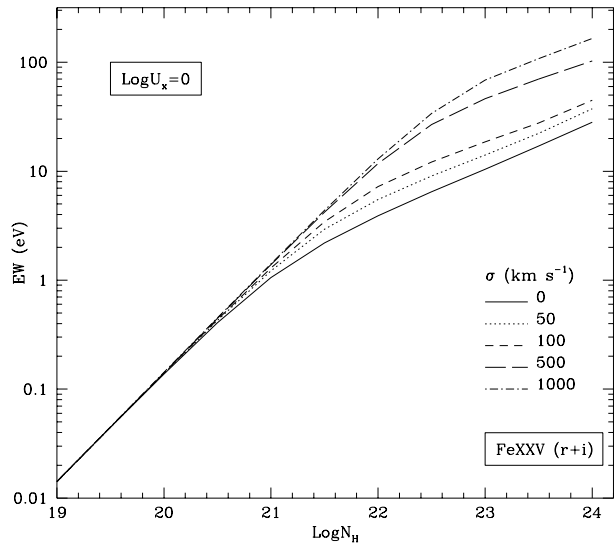}

\caption{\label{cogcsi25}Curve of growth for the Fe\,\textsc{xxv} K$\alpha$ absorption complex (resonant + intercombination) for different values of turbulence.}

\end{figure}

\begin{figure}

\epsfig{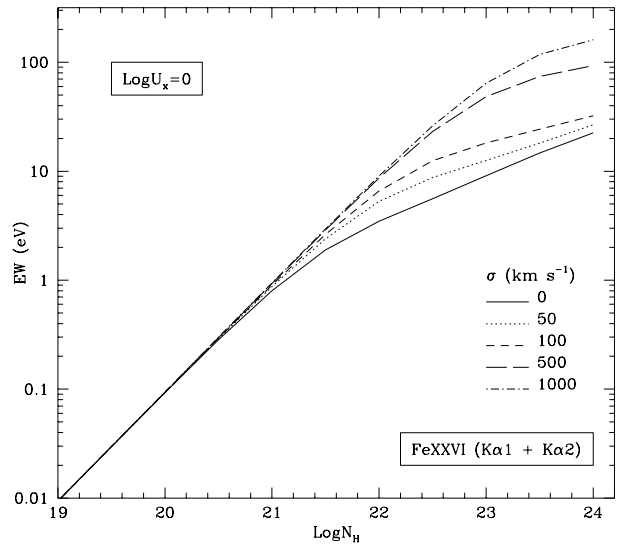}

\caption{\label{cogcsi26}Curve of growth for the Fe\,\textsc{xxvi} K$\alpha$ absorption complex (K$\alpha_1$ + K$\alpha_2$) for different values of turbulence.}

\end{figure}

Indeed, the effect of turbulence can be very important, because it broadens the line profile, effectively raising the column of gas needed for the line core to become optically thick. \citet{nfm99} have calculated that the optical depth to resonant absorption of a column of gas under strong turbulence (i.e. $\sigma_v$ larger than 300 km s$^{-1}$) can be more than 10 times lower than the same column of gas unaffected by turbulence. As a result, the curve of growth of the line saturates at a larger column density, allowing the EW to rise linearly over a wider range of $N_H$. This is shown in Fig. \ref{cogcsi25} and \ref{cogcsi26}: the resulting EWs can reach 100 eV for $N_H\simeq10^{24}$ cm$^{-2}$.

Finally, we explored the effects due to a different set of input parameters for the incident continuum, in terms of temperature, photon index and density. As found by \citet{bm02} for the emission lines, the results are basically the same for a wide range of gas densities, {because it turns out to be equivalent to change the distance between the source and the gas, while leaving the same ionisation parameter. On the other hand, an almost negligible variation in the position and/or the value of the maximum is found for plots \ref{fe25abs} and \ref{fe26abs} when different values of $\Gamma$ are explored, thanks to our choice of the X-ray ionisation parameter. A quantitatively more significant dependence on temperature is instead found, because of the larger contribution of collisional ionisation as $T$ increases, thus shifting to lower values the ionisation parameter corresponding to large abundances of \mbox{Fe\,\textsc{xxv}} and \mbox{Fe\,\textsc{xxvi}}. However, all these effects are similar to those presented by \citet{bm02} and therefore are not further discussed here.

\subsection{\label{emi}Emission lines}

Emission lines from \mbox{Fe\,\textsc{xxv}} and \mbox{Fe\,\textsc{xxvi}} in unobscured AGN were treated in detail by \citet{bm02}, to whom we refer the reader for details on calculations and for plots similar to the ones presented in this paper for the absorption features.

\mbox{Fe\,\textsc{xxv}} emission lines are produced by recombination
and resonant scattering \citep[see][ and references
therein]{bm02,mbf96}. Recombination lines in a photoionised matter are generally described by
introducing an effective fluorescent yield which represents the
probability that the recombination cascade includes the radiative
emission of a K$\alpha$ photon. This process can be treated
analytically in the Compton-thin regime, which is appropriate for
column densities up to $\simeq10^{24}$ cm$^{-2}$. As pointed out by
several authors \citep[see e.g.][]{pd00,bk00}, photoionised gas is
characterised by weak emission from the resonant \textit{w} line, while
for iron the two intercombination lines \textit{x}+\textit{y} contribute almost equally
with the forbidden \textit{z} line to the total emission feature, for a wide
range of densities. Therefore, if the lines are not resolved by the
X-ray instruments, a single feature with centroid around 6.66 keV
should come out from recombination processes in a photoionised gas.

Resonantly scattered K$\alpha$ lines are instead the result of
radiative de-excitation after excitation due to the absorption of
continuum photons at the energy of the line. As illustrated in Sect.
\ref{abs}, the dominant absorption transition is the resonant \textit{w},
followed by the intercombination \textit{y}. Since the latter has an oscillator
strength which is 10\% of the former, we expect that this ratio is
preserved in the emission process. The contribution from \textit{x} and \textit{z} is
completely negligible. Again, if the resolution of the X-ray detector
is lower than the separation between the \textit{w} and the \textit{y} line, we should
observe a blended feature around 6.70 keV. This process is much more
effective than recombination so that lines produced in this way are
dominant in the optically-thin regime. However, its cross-section is
much larger than in the previous case, so that the gas becomes rapidly
optically-thick to resonant scattering. By means of Monte Carlo
simulations, \citet{mbf96} showed that a column density as small as
$5\times10^{20}$ cm$^{-2}$ is enough to significantly reduce the EWs from these
lines. As a consequence, when $N_H$ becomes greater than
$\simeq10^{23}$ cm$^{-2}$, recombination becomes the dominant process.

Figure \ref{wxyz} summarizes these considerations, showing a plot of
the relative contributions of each transition as a function of the
column density of the gas. The plot takes into account the effect due
to the saturation of the resonant line when the column density
increases \citep[from figure 5 of][]{mbf96} and the discussion above
about the ratios between \textit{w}, \textit{x}, \textit{y} and
\textit{z} when produced by recombination or resonant scattering. As a
result, the resonant \textit{w} line dominates the emission spectrum up to
$10^{22}$ cm$^{-2}$ and then is reached and overcome by the forbidden
line \textit{z} and the two intercombination \textit{x} and \textit{y}. These results are in good agreement with the recent calculations performed by \citet{coupe04}.

Note that the total EW for the \mbox{Fe\,\textsc{xxv}} is the one
calculated by \citet{bm02}, while figure \ref{wxyz} simply contains
the information on how this EW should be divided among the different
lines. Therefore, an high resolution spectrum of the He-like iron
K$\alpha$ emission line complex produced in photoionised gas has, at
least in principle, the potentiality to reveal the column density of
the gas where the lines are produced, similarly to what \citet{kin02} already showed for lighter metals in the soft X-ray spectrum of NGC~1068. Moreover, it can also be useful
to note that, even if the resolution of the observed spectrum does not
allow to resolve the single lines, it could still be possible to
appreciate the difference between a 6.70 keV line at low column
densities and a 6.66 keV line at larger $N_H$.

Another interesting consequence of these results is the possibility to
observe simultaneously absorption and emission features from the same
ionic species. As an example, let us consider a large covering factor
gas, with high column density and lying \textit{also} along the line
of sight. In this case the resonant line \textit{w} would not be produced in
emission, but only absorbed, while \textit{x}, \textit{y} and \textit{z} lines would dominate the
emission He-like Fe complex. Therefore, the observed spectrum would
show the intercombination and forbidden lines in emission (the \textit{y} line
partly diluted by absorption), while the resonant line would be
apparent only in absorption: such an observation is in principle
within reach of \textit{Astro-E2}.

\begin{figure}

\epsfig{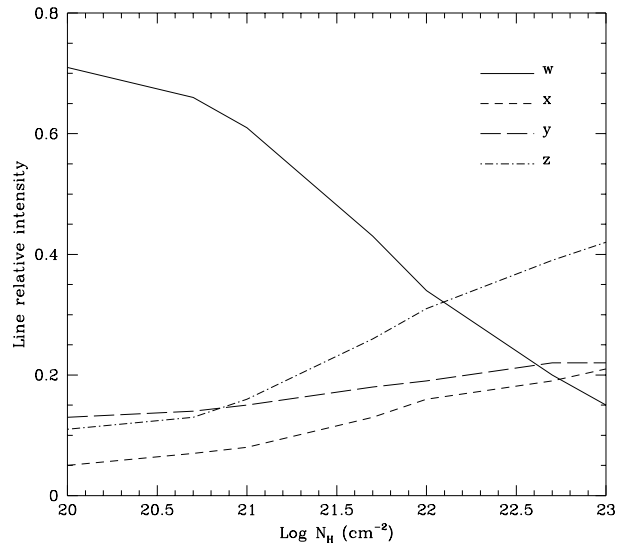}

\caption{\label{wxyz}Relative intensity of \mbox{Fe\,\textsc{xxv}} \textit{w}, \textit{x}, \textit{y} and \textit{z} transitions vs. the column density of the gas. Turbulent velocity is set equal to zero. Note that this is a rough plot: see text for details and caveats.}

\end{figure}

However, we would like to point out some cautionary warnings on these results. Figure \ref{wxyz} does not intend to be a detailed description of such a complex issue, but simply displays a rough behaviour deriving from the above-mentioned effects. The plot was produced in the transmission case, when the observer looks at the photons which come out from the cloud: in the reflection case, the dominance of resonant scattering at low column densities is enhanced, but the differences are not very significant for the observed spectrum \citep[see][ for details]{bm02,mbf96}.
It should also be reminded that inner-shell ionization of \mbox{Fe\,\textsc{xxiv}} is an additional mechanism in populating the upper level of the forbidden line (metastable level), hence if the ionic abundance of \mbox{Fe\,\textsc{xxiv}} is significant, this process can enhance the values of \textit{z} plotted in figure \ref{wxyz}. This can be the case in transient plasma as shown by \citet{ms75}.
Another contribution not considered in the plot is that from turbulence. As explained for the absorption lines in Sect. \ref{abs}, turbulence broadens the line profile, so that the gas becomes optically-thick to resonant scattering at a larger column density. As a consequence, the value of $N_H$ which separates the regimes where recombination or resonant scattering dominates increases with the amount of turbulence. As a final remark, note that we assumed that the gas was in pure photoionisation equilibrium: any deviation from this ideal situation could change significantly the ratio between the He-like emission lines.
 
\subsection{\label{satellite}Satellite lines}

 Satellite lines are systems of lines appearing close to or blended with the main lines of 
highly charged ionized atoms. 
These lines can either correspond to the stabilizing transitions in
the process of dielectronic recombination (dielectronic satellite lines) or by
 inner-shell electron/photon excitation (inner-shell excitation satellite lines).
As shown by \citet{op01} dielectronic satellite lines in hot plasmas (i.e. collisional plasmas) 
dominate the K$\alpha$ emission
 at temperature well below the one corresponding to the maximum of ionic abundance of FeXXV,
 i.e. T$<$ T$_{\rm m}$ which is approximatively 10$^{7.4}$\,K \citep{ar85}.

\begin{table}
\caption{\label{satlines}Experimental energies of the \mbox{Fe\,\textsc{xxv}} lines and of the satellite lines
 of \mbox{Fe\,\textsc{xxiv}} and \mbox{Fe\,\textsc{xxiii}} \citep[data from][]{dec97}.
 The oscillator strengths ($f_{\rm ij}$) greater than 0.05 are given: (a) from NIST (see Table 1),
 (b) from \citet{npz01}, and (c): from \citet{bn02}.}
\begin{tabular}{@{\ }l@{\ }l@{\ }c@{\ }c@{\ }c}
\hline
\hline
  Ion                     &  transition &  label     &  E & {\bf $f_{\rm ij}$} \\
                          &             &            & (keV) &   \\
\hline
\mbox{Fe\,\textsc{xxv}} &{\small  1s$^{2}$ $^{1}S_{0}$--1s\,2p $^{1}P_{1}$}                            & $w$     &  6.700 & 0.704$^{(a)}$  \\
\mbox{Fe\,\textsc{xxv}} & {\small  1s$^{2}$ $^{1}S_{0}$--1s\,2p $^{3}P_{2}$}                           & $x$     &  6.683 & \\
\mbox{Fe\,\textsc{xxiv}}& {\small  1s$^{2}$\,2s $^{2}S_{1/2}$--1s\,2s\,2p ($^{1}$P)\,$^{2}P_{1/2}$} & $t$     &  6.676 &0.110$^{(b)}$ \\
\mbox{Fe\,\textsc{xxv}} & {\small  1s$^{2}$ $^{1}S_{0}$--1s\,2p $^{3}P_{1}$}                           & $y$     &  6.668 & 0.069$^{(a)}$  \\
\mbox{Fe\,\textsc{xxiv}}& {\small  1s$^{2}$\,2s $^{2}S_{1/2}$--1s\,2s\,2p ($^{3}$P)\,$^{2}P_{3/2}$} & $q$     &  6.662 & 0.469$^{(c)}$  \\
\mbox{Fe\,\textsc{xxiv}}& {\small  1s$^{2}$\,2s $^{2}S_{1/2}$--1s\,2s\,2p ($^{3}$P)\,$^{2}P_{1/2}$} & $r$     &  6.653 & 0.147$^{(c)}$  \\
\mbox{Fe\,\textsc{xxv}} & {\small  1s$^{2}$ $^{1}S_{0}$--1s\,2p $^{3}S_{1}$}                           & $z$     &  6.637 &  \\
\mbox{Fe\,\textsc{xxiii}} & {\small  1s$^{2}$\,2s $^{2}$ $^{1}S_{0}$--1s\,2s$^{2}$\,2p\,$^{1}P_{1}$}       & $\beta$ &  6.629 & 0.666$^{(c)}$   \\
\hline
\end{tabular}
\end{table}

 Dielectronic recombination rates are very sensitive to temperature and become quickly
 negligible with respect to radiative recombination rates for temperatures around or below $10^6$ K,
 as appropriate for the photoionised gas under investigation in this paper
 \citep[see e.g.][ for the rates and the different temperature dependence of the two processes]{bd82,ar85}.
 As a result, contributions from dielectronic satellite lines can be neglected for our purposes. Inner-shell electron excitation also requires high temperature to be effective, so that satellite lines produced by this process are negligible.
 
On the contrary, the satellite lines to the \mbox{Fe\,\textsc{xxv}} lines ($w$, $x$, $y$, $z$) 
due to inner-shell photo-excitation of Li-like iron (\mbox{Fe\,\textsc{xxiv}})
 or Be-like iron (\mbox{Fe\,\textsc{xxiii}}) can be significant:\\

\noindent $1s^{2}\,2s + {\rm photon}   \rightarrow 1s\,2s\,2p $ ~~~ (satellite lines: $q$, $r$, $s$, $t$) \\

\noindent $1s^{2}\,2s^{2} + {\rm photon}   \rightarrow 1s\,2s^{2}\,2p $ ~~~ (satellite line: $\beta$) \\

\noindent Table~\ref{satlines} reports the experimental energies of the satellite lines in the vicinity of the
 \textit{w}, \textit{x}, \textit{y}, and \textit{z} lines \citep{dec97}. The values reported in Table~\ref{parameters} from the 
 database NIST are in excellent agreement with these experimental energies.
The oscillator strengths greater than 0.05 are given to show which lines are expected to be significant in
both absorption and emission. The two strongest satellite lines would be $q$ (Li-like) and $\beta$ (Be-like)
depending on the relative abundance of \mbox{Fe\,\textsc{xxiv}} and \mbox{Fe\,\textsc{xxiii}}, respectively.

 The spectral resolution of the XRS (X-Ray Spectrometer)
 on board \textit{ASTRO-E2} will be of the order of 6\,eV at 6.6-6.7 keV (\mbox{Fe\,\textsc{xxv}} lines).
 The lines $x$ and $t$ ($\Delta E_{x-t}$=7 eV), and the lines $z$ and $\beta$ ($\Delta E_{z-\beta}$=8 eV)
 can thus be resolved. 
 The lines $y$ and $q$ are separated by 6 eV which is close to the resolving power of the XRS, 
and it will be possible to infer the intensity of the line by line profile deconvolution. 
 We remind that the resolving power of \textit{Astro-E2} XRS is much larger than the thermal line width for iron at 10$^{6}$\,km\,s$^{-1}$ (well below 1 eV), while it corresponds to a turbulence velocity of about 270\,km\,s$^{-1}$.
Therefore, in case of significant line broadening due to turbulence, the contribution 
of the satellite lines to the four lines of \mbox{Fe\,\textsc{xxv}} has to be taken into account.

Finally, in very dense plasmas the 1s$^{2}$\,2p level can be populated and one can for example obtain: 
$1s^{2}\,2p + {\rm photon}  \rightarrow  1s\,2p^{2}$ (satellite lines: $k$ (6.654 keV), $j$ (6.644 keV)). 
However in this work we consider a plasma with a electron density of 10$^{6}$\,cm$^{-3}$ (Table~\ref{parameters}),
 and therefore the satellite lines $k$, and $j$ can be neglected here.
For \mbox{Fe\,\textsc{xxvi}}, no inner-shell excitation satellite line of \mbox{Fe\,\textsc{xxv}} exists.

In conclusion, in case of negligible turbulence velocity, i.e. less than about 270\,km\,s$^{-1}$,
 the satellite lines are unblended or are easily separated by line profile deconvolution to the four lines
 of \mbox{Fe\,\textsc{xxv}} and the lines intensities reported in this paper can be used safely for spectral analysis  
of {\textit{Astro-E2}}-XRS observations. 

\section{\label{observation}Observational evidence}

Tables \ref{cthick} and \ref{observation} show a complete (as far as we know) list of objects (Compton-thick and unobscured/Compton-thin, respectively) where emission and/or absorption lines from {Fe\,\textsc{xxv}} and {Fe\,\textsc{xxvi}} were detected by either \textit{Chandra} or XMM-\textit{Newton}. The listed values are those reported in the literature, as found in the cited references. When some of the needed values were not present in these works, we re-analysed the relative spectra and extracted the required information. Finally, for some objects we found in the archives XMM-\textit{Newton} and/or \textit{Chandra} unpublished data: we analysed them for the first time and included them in the table. Standard procedures for data reduction were adopted \citep[see e.g.][]{bianchi04}.

\subsection{Compton-thick objects}

Compton-thick Seyfert galaxies, i.e. objects obscured by a column density larger than $\sigma_\mathrm{T}^{-1}=1.5\times10^{24}$ cm$^{-2}$, are the best suited objects for the observation of ionised iron lines in emission, since the primary continuum is completely suppressed in this band and hence these lines are not diluted by this component. We present in Table \ref{cthick} \textit{Chandra} and XMM-\textit{Newton} results from three of the brightest Compton-thick Seyfert galaxies, that is NGC~1068, Circinus and Mrk~3.

\begin{table*}
\caption{\label{cthick}{Fe\,\textsc{xxv}} and {Fe\,\textsc{xxvi}} emission lines in three of the brightest Compton-thick Seyfert galaxies. A reference to this paper (\textit{TP}) means that data were re-analysed to add some values not found in literature (in these cases, numbers are indicated in italics)}

\begin{tabular}{|c|c|c|c|c|c|c|}
\hline Source & Instr. [Date]& \multicolumn{2}{c}{\mbox{Fe\,\textsc{xxv}}} & \multicolumn{2}{c}{\mbox{Fe\,\textsc{xxvi}}} & Ref. \\
& [mm/yy] & EW$^a$ (eV) & I$^b$ & EW$^a$ (eV) & I$^b$ &  \\
\hline \multirow{3}*{\textbf{NGC~1068}} & \multirow{2}*{EPIC pn [07/00]} & \textit{w} ($6.725\pm0.001$ keV): $1430^{+195}_{-130}$ & $2.2^{+0.3}_{-0.2}$ & \multirow{2}*{$494^{+210}_{-70}$} & \multirow{2}*{$0.7^{+0.3}_{-0.1}$} &\multirow{2}*{\citet{matt04}}\\
& &\textit{z} ($6.61^{+0.01}_{-0.04}$ keV): $485^{+120}_{-60}$ & $0.8^{+0.2}_{-0.1}$ & & &\\
&\multirow{2}*{ACIS HETG [02/00]}& \textit{w} ($6.73\pm0.02$ keV): $\mathit{1000\pm500}$ & $1.5\pm0.7$ & \multirow{2}*{$\mathit{<800}$} & \multirow{2}*{$<1.1$} & \multirow{2}*{\citet{ogle03}, \textit{TP}}\\
& &\textit{z} ($6.59\pm0.02$ keV): $\mathit{1000\pm500}$ & $1.6\pm0.7$ & & &\\
\hline
\multirow{2}*{\textbf{Circinus}} & EPIC pn [08/01] & $\mathit{420\pm60}$ & $1.4\pm0.2$ & $\mathit{250^{+160}_{-130}}$ & $\mathit{0.8^{+0.5}_{-0.4}}$ & \citet{mbm03}, \textit{TP}\\
&ACIS HETG [06/00] & $\mathit{800\pm330}$ & $2.7\pm1.1$ & $\mathit{<310}$ & $\mathit{<2.6}$& \citet{sambruna01b}\\
\hline
\multirow{2}*{\textbf{Mrk~3}} & EPIC pn [10/00] & $4800\pm2400$& $0.4\pm0.2$ & $<1500$& $<0.1$ & Bianchi et al., in prep.\\
& ACIS HETG [03/00] & $\mathit{11\,000^{+9000}_{-4000}}$  & $\mathit{0.9^{+0.7}_{-0.3}}$ & $\mathit{<10\,000}$ & $\mathit{<2.0}$ & \citet{sako00b}, \textit{TP}\\
\hline
\end{tabular}

$^a$ Calculated against the reflected continuum only -- $^b$ $10^{-5}$ photons cm$^{-2}$ s$^{-1}$

\end{table*}

The most interesting object is NGC~1068, where the resonant \textit{w} and the forbidden \textit{z} lines are actually resolved in the XMM-\textit{Newton} data. The ratio of their fluxes, $F_w/F_z=2.8\pm0.8$, points towards a gas column density range roughly of $\log{N_H}=21\div21.5$ (see Fig. \ref{wxyz}). On the other hand, an ionization parameter around $\log{U_x}=-0.5$ can be measured from the ratio between the {Fe\,\textsc{xxv}} and {Fe\,\textsc{xxvi}} EWs, thus requiring an iron overabundance of a factor $\simeq2$ with respect to the solar values to obtain the total EW of each line} \citep[see Fig. 2, 3, 4, 5 from][]{bm02}. Note that such an overabundance was also measured from the depth of the iron edge in the EPIC pn data \citep{matt04}: even if these values correspond to two different materials, it is suggestive that they are similar.

A much lower flux ratio between the the resonant \textit{w} and the forbidden \textit{z} lines is instead measured in the \textit{Chandra}
data. Even if this value is still consistent within the errors with
that found in the XMM-\textit{Newton} data and the two lines are not
resolved in the HETG spectrum, being actually the result of a
de-blending of a single feature \citep[see][]{ogle03}, it may still
give some indications on the geometry of the photoionised reflector.
In particular, \citet{ogle03} noted that this region should be
extended on a 3 arcsec scale, implying that the HETG spectrum includes
only the inner region, while the EPIC pn is clearly observing
all the gas. In this case, it is possible that the gas which is closer
to the nucleus has a larger column density, enhancing the production
of the forbidden line, while the outer regions, which dominate the pn
spectrum, have a lower column density. However, the nuclear region in NGC~1068 accounts for the major part of the flux \citep{yws01,brink02,ogle03}, so that any contribution from the extended region to the total EPIC pn spectrum should be small. Moreover, this scenario is at odds with
the fact that the total fluxes of the {Fe\,\textsc{xxv}} features seem
constant between the two datasets, but the very large errors on the
\textit{Chandra} flux values prevent us from drawing any firm
conclusion.

As a final remark on NGC~1068, we note that the centroid energies for the {Fe\,\textsc{xxv}} lines are formally inconsistent with their expected values, both in the EPIC pn and in the \textit{Chandra} spectra. This is puzzling, because in both observations a blueshift for the \textit{w} line and a redshift for the \textit{z} line are implied, at odds with a common origin in the same material. Moreover, the {Fe\,\textsc{xxvi}} line energy \citep[$6.92^{+0.01}_{-0.04}$ keV:][]{matt04} also implies a redshift. While a less than perfect calibration can be invoked for both instruments, we do not have any other satisfactory explanation for this result.

The other two Compton-thick sources have ionised iron lines much
weaker with respect to the total continuum and to the neutral iron
line, so that the statistics do not allow us to resolve the single
transitions. However, in the case of Mrk~3, a best fit value of
$6.71^{+0.03}_{-0.02}$ keV for the {Fe\,\textsc{xxv}} line (Bianchi et al., in
preparation) seems to indicate that the resonant line is the dominant
one also in this source, again suggesting a low column density. In
both sources, the He-like line is much stronger than the H-like, so
that the ionisation parameter should not exceed $\log{U_X}\simeq-0.5$.
From the observed EWs, an iron overabundance of at least a factor 2 is
needed for Mrk~3, while a value closer to 1 is required for Circinus,
as again found from the depth of the iron edge \citep{mbm03}.

Finally, it must be mentioned that there are two other Compton-thick
sources bright enough to allow a detailed spectral analysis, i.e.
NGC~6240 and NGC~4945 \citep{matt00c}. In the former source both the
He- and H-like iron lines are detected \citep{boller03}, while in the
latter only the He-like is visible \citep{schurch02}. However, as
argued in the abovementioned papers, it is possible that in these two
sources, which have strong starburst regions, the lines are emitted in
collisionally ionised plasma, and therefore we do not discuss them
here.

\subsection{Unobscured and Compton-thin objects}

 The sample in Table \ref{observation} includes a majority of Seyfert 1s, some intermediate objects and a Compton-thin Seyfert 2 (NGC~4507). As for the X-ray properties, all the sources are either unobscured or the absorbing column density from neutral matter allows the direct observation of the nuclear continuum at the iron line K band. 

\begin{table*}
\caption{\label{observation}A list of objects where emission and/or absorption lines from {Fe\,\textsc{xxv}} and {Fe\,\textsc{xxvi}} were detected (see text for details). A reference to this paper (\textit{TP}) means that data were re-analysed to add some values not found in literature (in these cases, numbers are indicated in italics) or that the relative observation was analysed for the first time.}

\begin{tabular}{|c|c|c|c|c|c|c|}
\hline Source & Instr. [Date]& \multicolumn{2}{c}{\mbox{Fe\,\textsc{xxv}}} & \multicolumn{2}{c}{\mbox{Fe\,\textsc{xxvi}}} & Ref. \\
& [mm/yy] & EW (eV) & I$^a$ & EW (eV) & I$^a$ &  \\
\hline
\hline \multicolumn{7}{c}{\textsc{Emission lines}}\\
\hline \multirow{3}*{\textbf{NGC~5506}}&ACIS HETG [12/00] & $\mathit{<31}$& $\mathit{<2.4}$&  $\mathit{<26}$ & $\mathit{<1.7}$ &\multirow{3}*{\citet{bianchi03}, \textit{TP}}\\
&EPIC pn [02/01] &$27^{+17}_{-12}$  & $2.0^{+1.3}_{-0.9}$ & $41^{+11}_{-27}$&$2.0^{+0.5}_{-1.3}$ &\\
&EPIC pn [01/02]& $\mathit{19^{+11}_{-12}}$ & $\mathit{2.3^{+1.5}_{-1.4}}$ & $\mathit{19\pm13}$ & $\mathit{2.0^{+1.5}_{-1.4}}$ &\\
\hline \textbf{NGC~7213} & EPIC pn [05/01]&$25^{+10}_{-13}$ & $0.6\pm0.3$ &$22\pm14$& $0.5\pm0.3$ &\citet{bianchi03b}\\
\hline \multirow{2}*{\textbf{NGC~7314}}& EPIC pn [05/01]&$\mathit{16\pm11}$ & $\mathit{0.7\pm0.5}$ & $\mathit{28\pm13}$& $\mathit{1.1\pm0.5}$ &\textit{TP} \\
& ACIS HETG [07/02]& $29^{+20}_{-17}$ & $1.1^{+0.7}_{-0.6}$ & $63^{+33}_{-23}$ & $1.9^{+1.0}_{-0.7}$ & \citet{yaq03} \\
\hline \multirow{2}*{\textbf{IC~4329A}}& EPIC pn [01/01]& $\mathit{<13}$ & $\mathit{<2.1}$ & $\mathit{<48}$ & $\mathit{<6.3}$ & \textit{TP}\\
&ACIS HETG [08/01]& $<17$ & $\mathit{<4.6}$ & $42^{+30}_{-21}$ & $6.2^{+4.4}_{-3.1} $ &\citet{my04}, \textit{TP}\\
\hline \multirow{2}*{\textbf{NGC4593}} &ACIS HETG [06/01]& $\mathit{<31}$ & $\mathit{<1.5}$ &$\mathit{<72}$ & $\mathit{<2.7}$ & \textit{TP}\\
& EPIC pn [06/02]& $<13$ & $\mathit{<1.0}$ & $39\pm13$ & $\mathit{1.4^{+1.1}_{-0.6}}$ & \citet{rey04}, \textit{TP}\\
\hline \textbf{MCG-2-58-22} &EPIC pn [12/00]&$\mathit{<44}$ & $\mathit{<1.6}$ & $\mathit{55^{+26}_{-35}}$ &$\mathit{1.8^{+0.8}_{-1.1}} $ & \textit{TP}\\
\hline \textbf{MRK205} & EPIC pn [05/00]& $35\pm17$ & $\mathit{0.2\pm0.1}$ & $51\pm19$ & $\mathit{0.3\pm0.1}$ & \citet{reev01}, \textit{TP}\\
\hline \textbf{NGC~3783} & EPIC pn [12/01]& --$^b$ & --$^b$ &$17\pm5$ & $1.4\pm0.3$ & \citet{reev04}\\
\hline \multirow{3}*{\textbf{ESO198-G024}} & EPIC pn [12/00] & $\mathit{<80}$ & $\mathit{<0.9}$ & $\mathit{<65}$ & $\mathit{<1.0}$ &\multirow{3}*{\textit{TP}}\\
&EPIC pn [01/01] & $\mathit{<28}$ & $\mathit{<0.4}$ & $\mathit{<63}$ & $\mathit{<0.8}$ &\\
& ACIS-S [04/04]& $\mathit{<15}$ & $\mathit{<0.1}$ & $\mathit{56^{+49}_{-35}}$ & $\mathit{0.3\pm0.2}$ & \\
\hline
\hline \multicolumn{7}{c}{\textsc{Absorption lines}}\\
\hline \multirow{2}*{\textbf{NGC~4507}}& EPIC pn [01/01]& $\mathit{>-16}$ & --$^c$ & $\mathit{>-24}$ & --$^c$ &\multirow{2}*{\citet{matt04b}, \textit{TP}} \\
& ACIS HETG [03/01]& $-26\pm16$ & --$^c$ & $\mathit{>-12}$ & --$^c$ &  \\
\hline \textbf{MCG-6-30-15} & EPIC pn [07/00+07/01]& $-11\pm6$ & --$^c$ & --$^d$ & --$^c$ & \citet{vf04} \\
\hline \multirow{2}*{\textbf{NGC~3783}} & EPIC pn [12/01] & $-17\pm5$ & --$^c$ & --$^b$ & --$^c$ & \citet{reev04}\\
&ACIS HETG [01/00+06/01]& $-13\pm5$ & --$^c$ &  --$^b$ & --$^c$ & \citet{kaspi02} \\
\hline

\end{tabular}

$^a$ $10^{-5}$ photons cm$^{-2}$ s$^{-1}$ -- $^b$ Fluxes are not considered for variability in absorption lines (see text for details) -- $^c$ Values are not calculated because a line in emission (absorption) of the same ionic species is present in the spectrum -- $^d$ A detailed analysis of this complex spectrum, which includes EPIC pn and MOS data from two observations, is beyond the scopes of this paper.

\end{table*}

All three sources with absorption lines present only a feature from He-like iron, thus implying $\log U_x<0$ (figure \ref{ratio}). The observed EWs can be produced by column densities around $10^{23}$ cm$^{-2}$, except for NGC~4507 which needs a larger $N_H$, unless effects from turbulence and/or iron abundance are taken into account.

On the other hand, it is interesting to note that 4 out of the nine sources where emission lines are observed have indeed features both from \mbox{Fe\,\textsc{xxv}} and \mbox{Fe\,\textsc{xxvi}}. In these cases, an ionisation parameter around $\log U_x=0\div0.5$ is required. The observed EWs are so large to imply column densities larger than $10^{23}$ cm$^{-2}$ and/or iron overabundance by a factor 2-3. This is also true for the remaining objects, where only the H-like line is observed, even if an higher ionisation parameter (around $\log U_x=0.5$) is appropriate in these cases. However, it is important to stress that the errors on the EWs are generally so large to possibly reduce the need for iron overabundance in many cases.

NGC~3783 is the only object whose spectrum displays both an absorption and an emission line. These lines are from different ionic species, respectively from \mbox{Fe\,\textsc{xxv}} and \mbox{Fe\,\textsc{xxvi}}. Since one is seen in absorption and the other in emission, they must be produced at least in two different clouds, the former being along the line of sight. If this is the case, the two clouds may well have different values of the ionisation parameter, because of a different value of the density or of the distance from source. Under these conditions, column densities of few $10^{22}$ cm$^{-2}$ and $\log U_x \simeq -0.5\div0$ can be responsible for the observed EW of the cloud seen in emission, while $N_H$ larger by an order of magnitude and $\log U_x \simeq 0\div0.5$ are required for the one in absorption. No particular need for overabundance or other corrections are needed.

Unfortunately, the quality of present spectra does not allow us to take advantage of the He-like diagnostics presented in the previous section, because the energy resolution is not good enough to disentangle the contributions from the single transitions. On the other hand, the effective area of the instruments does not provide enough statistics to measure the centroid energy of these weak lines with a precision sufficient to check for a clear difference between a recombination or resonant scattering dominated regime.

The larger occurrence of emission lines with respect to absorption lines in Table \ref{observation} may be giving us some information on the covering factor $f_c$ of the photoionised material, i.e. the solid angle subtended by the gas under investigation. Indeed, a prevalence of emission lines with large EWs is expected for large $f_c$, even if this relation is not linear for high column densities and when additional effects are taken into account, so that the maximum emission is typically for $f_c \simeq 0.5-0.7$ \citep{netzer93,netzer96}. On the other hand, absorption lines are dominant when a small covering factor cloud ($f_c<0.1$) lies exactly along the line of sight, so that the emission lines from other directions gives a negligible contribution to the overall spectrum, as in the case adopted by \citet{nfm99}. Therefore, even if it is likely that in low quality spectra emission lines are easier to observe than absorption features, we can conclude that the gas we are studying generally subtends a considerable solid angle with respect to the illuminating source.

The flux variability of the observed emission lines could be in principle another tool to investigate the geometrical properties of the gas that produces them. Indeed, a variation of the line flux would imply a response of the reflected spectrum to a variation of the illuminating source, with a delay proportional to the distance of the reflecting medium. In all the cases where we have more than one observation, no variation in the fluxes of the lines are detected. Even if this result is not very strong due to the large errors in the data, it is still an indication that the region responsible for the production of the lines from \mbox{Fe\,\textsc{xxv}} and \mbox{Fe\,\textsc{xxvi}} should be located at a distance at least of the order of light-months, up to  years in many cases, from the illuminating source.

As for the absorption lines, the relevant parameter for variability is not the flux, but the EW. In NGC~3783, \citet{reev04} reported a significant EW variability within the same observation, implying a change in the properties of the gas in a timescale smaller than $10^5$ s. However, the integrated EW of the six \textit{Chandra} observations analysed by \citet{kaspi02} is still consistent with that measured in the total EPIC pn spectrum and no variability was found in the soft X-ray spectrum of this source \citep{beh03,netz03}. On the other hand, \citet{vf04} reported an EW variability in the absorption line of MCG-6-30-15 between the two EPIC observations taken a year apart. The interpretation of such variations is less straightforward than for the emission lines, because it may be in principle due either by a change in the ionisation structure of the gas or in its column density along the line of sight.

\section{Conclusions}

We have calculated the equivalent widths (EWs) of the absorption lines produced by {Fe\,\textsc{xxv}} and {Fe\,\textsc{xxvi}} in a Compton-thin, low-velocity photoionised material illuminated by the nuclear continuum in AGN. We found that our results are consistent with the observations, provided column densities of the gas around $10^{23}$ cm$^{-2}$ or larger. However, we have shown that these values can be effectively reduced if the effects due to turbulence are taken in account.

On the other hand, the comparison of similar calculations for the emission lines \citep[as presented by][]{bm02} with an up-to-date list of known unobscured (at least at the iron line) Seyfert galaxies generally results in the need for an iron overabundance by a factor 2-3. However, it should be stressed that the error bars on the observed EWs are still too large to reach a firm conclusion. The much larger occurrence of emission lines with respect to the absorption ones in the analysed spectra gives an hint that the covering factor of the photoionised gas under investigation is likely quite large.

Finally, we have shown that the relative contributions of the single transitions of {Fe\,\textsc{xxv} \textit{w}, \textit{x}, \textit{y} and \textit{z} in the regimes where recombination or resonant scattering dominates provides a useful diagnostic tool to measure the column density of the gas. This effect must be taken in account with future, high resolution missions, like \textit{Astro-E2}. In the meantime, we have shown that in the high quality EPIC pn spectrum of the Compton-thick Seyfert galaxy NGC~1068, the forbidden \textit{z} and the resonant \textit{w} lines are actually resolved and their flux ratio implies a column density range of $\log{N_H}=21\div21.5$, in excellent agreement with what expected from the total EW.

\section*{Acknowledgements}

SB and GM acknowledge ASI and MIUR (under grant \textsc{cofin-03-02-23}) for financial support.

\bibliographystyle{mn}
\bibliography{sbs}

\begin{thebibliography}{45}
\expandafter\ifx\csname natexlab\endcsname\relax\def\natexlab#1{#1}\fi

\bibitem[{{Anders} \& {Grevesse}(1989)}]{ag89}
{Anders} E., {Grevesse} N., 1989, \gca, 53, 197

\bibitem[{{Antonucci}(1993)}]{antonucci93}
{Antonucci} R., 1993, \araa, 31, 473

\bibitem[{{Arnaud} \& {Rothenflug}(1985)}]{ar85}
{Arnaud} M., {Rothenflug} R., 1985, \aaps, 60, 425

\bibitem[{{Bautista} \& {Kallman}(2000)}]{bk00}
{Bautista} M.~A., {Kallman} T.~R., 2000, \apj, 544, 581

\bibitem[{{Behar} \& {Netzer}(2002)}]{bn02}
{Behar} E., {Netzer} H., 2002, \apj, 570, 165

\bibitem[{{Behar} {et~al.}(2003){Behar}, {Rasmussen}, {Blustin}, {Sako},
  {Kahn}, {Kaastra}, {Branduardi-Raymont}, \& {Steenbrugge}}]{beh03}
{Behar} E., {Rasmussen} A.~P., {Blustin} A.~J., {Sako} M., {Kahn} S.~M.,
  {Kaastra} J.~S., {Branduardi-Raymont} G., {Steenbrugge} K.~C., 2003, \apj,
  598, 232

\bibitem[{{Bely-Dubau} {et~al.}(1982){Bely-Dubau}, {Faucher}, {Dubau}, \&
  {Gabriel}}]{bd82}
{Bely-Dubau} F., {Faucher} P., {Dubau} J., {Gabriel} A.~H., 1982, \mnras, 198,
  239

\bibitem[{{Bianchi} {et~al.}(2003{\natexlab{a}}){Bianchi}, {Balestra}, {Matt},
  {Guainazzi}, \& {Perola}}]{bianchi03}
{Bianchi} S., {Balestra} I., {Matt} G., {Guainazzi} M., {Perola} G.~C.,
  2003{\natexlab{a}}, \aap, 402, 141

\bibitem[{{Bianchi} \& {Matt}(2002)}]{bm02}
{Bianchi} S., {Matt} G., 2002, \aap, 387, 76

\bibitem[{{Bianchi} {et~al.}(2004){Bianchi}, {Matt}, {Balestra}, {Guainazzi},
  \& {Perola}}]{bianchi04}
{Bianchi} S., {Matt} G., {Balestra} I., {Guainazzi} M., {Perola} G.~C., 2004,
  \aap, 422, 65

\bibitem[{{Bianchi} {et~al.}(2003{\natexlab{b}}){Bianchi}, {Matt}, {Balestra},
  \& {Perola}}]{bianchi03b}
{Bianchi} S., {Matt} G., {Balestra} I., {Perola} G.~C., 2003{\natexlab{b}},
  \aap, 407, L21

\bibitem[{{Boller} {et~al.}(2003){Boller}, {Keil}, {Hasinger}, {Costantini},
  {Fujimoto}, {Anabuki}, {Lehmann}, \& {Gallo}}]{boller03}
{Boller} T., {Keil} R., {Hasinger} G., {Costantini} E., {Fujimoto} R.,
  {Anabuki} N., {Lehmann} I., {Gallo} L., 2003, \aap, 411, 63

\bibitem[{{Brinkman} {et~al.}(2002){Brinkman}, {Kaastra}, {van der Meer},
  {Kinkhabwala}, {Behar}, {Kahn}, {Paerels}, \& {Sako}}]{brink02}
{Brinkman} A.~C., {Kaastra} J.~S., {van der Meer} R.~L.~J., {Kinkhabwala} A.,
  {Behar} E., {Kahn} S.~M., {Paerels} F.~B.~S., {Sako} M., 2002, \aap, 396, 761

\bibitem[{{Coup{\' e}} {et~al.}(2004){Coup{\' e}}, {Godet}, {Dumont}, \&
  {Collin}}]{coupe04}
{Coup{\' e}} S., {Godet} O., {Dumont} A.-M., {Collin} S., 2004, \aap, 414, 979

\bibitem[{{Decaux} {et~al.}(1997){Decaux}, {Beiersdorfer}, {Kahn}, \&
  {Jacobs}}]{dec97}
{Decaux} V., {Beiersdorfer} P., {Kahn} S.~M., {Jacobs} V.~L., 1997, \apj, 482,
  1076

\bibitem[{{Ferland}(2000)}]{ferl00}
{Ferland} G.~J., 2000, in Revista Mexicana de Astronomia y Astrofisica
  Conference Series, pp. 153--157

\bibitem[{{Kaspi} {et~al.}(2002){Kaspi}, {Brandt}, {George}, {Netzer},
  {Crenshaw}, {Gabel}, {Hamann}, {Kaiser}, {Koratkar}, {Kraemer}, {Kriss},
  {Mathur}, {Mushotzky}, {Nandra}, {Peterson}, {Shields}, {Turner}, \&
  {Zheng}}]{kaspi02}
{Kaspi} S., et al., 2002, \apj, 574, 643

\bibitem[{{Kinkhabwala} {et~al.}(2002){Kinkhabwala}, {Sako}, {Behar}, {Kahn},
  {Paerels}, {Brinkman}, {Kaastra}, {Gu}, \& {Liedahl}}]{kin02}
{Kinkhabwala} A., et al., 2002, \apj, 575, 732

\bibitem[{{Matt} {et~al.}(2004{\natexlab{a}}){Matt}, {Bianchi}, {D'Ammando}, \&
  {Martocchia}}]{matt04b}
{Matt} G., {Bianchi} S., {D'Ammando} F., {Martocchia} A., 2004{\natexlab{a}},
  \aap, 421, 473

\bibitem[{{Matt} {et~al.}(2004{\natexlab{b}}){Matt}, {Bianchi}, {Guainazzi}, \&
  {Molendi}}]{matt04}
{Matt} G., {Bianchi} S., {Guainazzi} M., {Molendi} S., 2004{\natexlab{b}},
  \aap, 414, 155

\bibitem[{{Matt} {et~al.}(1996){Matt}, {Brandt}, \& {Fabian}}]{mbf96}
{Matt} G., {Brandt} W.~N., {Fabian} A.~C., 1996, \mnras, 280, 823

\bibitem[{{Matt} {et~al.}(2000){Matt}, {Fabian}, {Guainazzi}, {Iwasawa},
  {Bassani}, \& {Malaguti}}]{matt00c}
{Matt} G., {Fabian} A.~C., {Guainazzi} M., {Iwasawa} K., {Bassani} L.,
  {Malaguti} G., 2000, \mnras, 318, 173

\bibitem[{{McKernan} \& {Yaqoob}(2004)}]{my04}
{McKernan} B., {Yaqoob} T., 2004, \apj, 608, 157

\bibitem[{{Mewe} \& {Schrijver}(1975)}]{ms75}
{Mewe} R., {Schrijver} J., 1975, \apss, 38, 345

\bibitem[{{Molendi} {et~al.}(2003){Molendi}, {Bianchi}, \& {Matt}}]{mbm03}
{Molendi} S., {Bianchi} S., {Matt} G., 2003, \mnras, 343, L1

\bibitem[{{Nahar} {et~al.}(2001){Nahar}, {Pradhan}, \& {Zhang}}]{npz01}
{Nahar} S.~N., {Pradhan} A.~K., {Zhang} H.~L., 2001, \pra, 63, 060701

\bibitem[{{Netzer}(1993)}]{netzer93}
{Netzer} H., 1993, \apj, 411, 594

\bibitem[{{Netzer}(1996)}]{netzer96}
---, 1996, \apj, 473, 781

\bibitem[{{Netzer} {et~al.}(2003){Netzer}, {Kaspi}, {Behar}, {Brandt},
  {Chelouche}, {George}, {Crenshaw}, {Gabel}, {Hamann}, {Kraemer}, {Kriss},
  {Nandra}, {Peterson}, {Shields}, \& {Turner}}]{netz03}
{Netzer} H., et al., 2003, \apj, 599, 933

\bibitem[{{Nicastro} {et~al.}(1999){Nicastro}, {Fiore}, \& {Matt}}]{nfm99}
{Nicastro} F., {Fiore} F., {Matt} G., 1999, \apj, 517, 108

\bibitem[{{Oelgoetz} \& {Pradhan}(2001)}]{op01}
{Oelgoetz} J., {Pradhan} A.~K., 2001, \mnras, 327, L42

\bibitem[{{Ogle} {et~al.}(2003){Ogle}, {Brookings}, {Canizares}, {Lee}, \&
  {Marshall}}]{ogle03}
{Ogle} P.~M., {Brookings} T., {Canizares} C.~R., {Lee} J.~C., {Marshall} H.~L.,
  2003, \aap, 402, 849

\bibitem[{{Porquet} \& {Dubau}(2000)}]{pd00}
{Porquet} D., {Dubau} J., 2000, \aaps, 143, 495

\bibitem[{{Pounds} {et~al.}(2003){Pounds}, {Reeves}, {Page}, {Wynn}, \&
  {O'Brien}}]{pounds03b}
{Pounds} K.~A., {Reeves} J.~N., {Page} K.~L., {Wynn} G.~A., {O'Brien} P.~T.,
  2003, \mnras, 342, 1147

\bibitem[{{Reeves} {et~al.}(2004){Reeves}, {Nandra}, {George}, {Pounds},
  {Turner}, \& {Yaqoob}}]{reev04}
{Reeves} J.~N., {Nandra} K., {George} I.~M., {Pounds} K.~A., {Turner} T.~J.,
  {Yaqoob} T., 2004, \apj, 602, 648

\bibitem[{{Reeves} {et~al.}(2003){Reeves}, {O'Brien}, \& {Ward}}]{reev03}
{Reeves} J.~N., {O'Brien} P.~T., {Ward} M.~J., 2003, \apjl, 593, L65

\bibitem[{{Reeves} {et~al.}(2001){Reeves}, {Turner}, {Pounds}, {O'Brien},
  {Boller}, {Ferrando}, {Kendziorra}, \& {Vercellone}}]{reev01}
{Reeves} J.~N., {Turner} M.~J.~L., {Pounds} K.~A., {O'Brien} P.~T., {Boller}
  T., {Ferrando} P., {Kendziorra} E., {Vercellone} S., 2001, \aap, 365, L134

\bibitem[{{Reynolds} {et~al.}(2004){Reynolds}, {Brenneman}, {Wilms}, \&
  {Kaiser}}]{rey04}
{Reynolds} C.~S., {Brenneman} L.~W., {Wilms} J., {Kaiser} M.~E., 2004, \mnras,
  352, 205

\bibitem[{{Sako} {et~al.}(2000){Sako}, {Kahn}, {Paerels}, \&
  {Liedahl}}]{sako00b}
{Sako} M., {Kahn} S.~M., {Paerels} F., {Liedahl} D.~A., 2000, \apjl, 543, L115

\bibitem[{Sambruna {et~al.}(2001)Sambruna, Netzer, Kaspi, Brandt, Chartas,
  Garmire, Nousek, \& Weaver}]{sambruna01b}
Sambruna R.~M., Netzer H., Kaspi S., Brandt W.~N., Chartas G., Garmire G.~P.,
  Nousek J.~A., Weaver K.~A., 2001, \apj, 546, L13

\bibitem[{{Schurch} {et~al.}(2002){Schurch}, {Roberts}, \&
  {Warwick}}]{schurch02}
{Schurch} N.~J., {Roberts} T.~P., {Warwick} R.~S., 2002, \mnras, 335, 241

\bibitem[{{Vaughan} \& {Fabian}(2004)}]{vf04}
{Vaughan} S., {Fabian} A.~C., 2004, \mnras, 348, 1415

\bibitem[{{Verner} {et~al.}(1996){Verner}, {Verner}, \& {Ferland}}]{vvf96}
{Verner} D.~A., {Verner} E.~M., {Ferland} G.~J., 1996, Atomic Data and Nuclear
  Data Tables, 64, 1

\bibitem[{{Yaqoob} {et~al.}(2003){Yaqoob}, {George}, {Kallman}, {Padmanabhan},
  {Weaver}, \& {Turner}}]{yaq03}
{Yaqoob} T., {George} I.~M., {Kallman} T.~R., {Padmanabhan} U., {Weaver} K.~A.,
  {Turner} T.~J., 2003, \apj, 596, 85

\bibitem[{{Young} {et~al.}(2001){Young}, {Wilson}, \& {Shopbell}}]{yws01}
{Young} A.~J., {Wilson} A.~S., {Shopbell} P.~L., 2001, \apj, 556, 6

\end{thebibliography}

\label{lastpage}

\end{document}